# Direct Experimental Evidence for Differing Reactivity Alterations of Minerals following Irradiation: The Case of Calcite and Quartz

*Isabella Pignatelli ([\*]), Aditya Kumar ([†]), Kevin G. Field ([‡]), Bu Wang ([§]), Yingtian Yu ([\*\*]), Yann Le Pape ([††]), Mathieu Bauchy ([‡‡]), Gaurav Sant ([§§],[\*\*\*])*

**Abstract**

Concrete, a mixture formed by mixing cement, water, and fine and coarse mineral aggregates is used in the construction of nuclear power plants (NPPs), e.g., to construct the reactor cavity concrete that encases the reactor pressure vessel, etc. In such environments, concrete may be exposed to radiation (e.g., neutrons) emanating from the reactor core. Until recently, concrete has been assumed relatively immune to radiation exposure. Direct evidence acquired on $Ar^+$-ion irradiated calcite and quartz indicates, on the contrary, that, such minerals, which constitute aggregates in concrete, may be significantly altered by irradiation. Specifically, while quartz undergoes disordering of its atomic structure resulting in a near complete lack of periodicity, i.e., similar to glassy silica, calcite only experiences random rotations, and distortions of its carbonate groups. As a result, irradiated quartz shows a reduction in density of around 15%, and an increase in chemical reactivity, described by its dissolution rate, similar to a glassy silica; i.e., an increase of ≈3 orders of magnitude. Calcite however, shows little change in dissolution rates – although its density noted to reduce by ≈9%. These differences are correlated with the nature of bonds in these minerals, i.e., being dominantly ionic or covalent, and the rigidity of the mineral's atomic network that is characterized by the number of topological constraints ($n_c$) that are imposed on the atoms in the network. The outcomes are discussed within the context of the durability of concrete structural elements formed with calcitic/quartzitic aggregates in nuclear power plants.

---

[\*] Laboratory for the Chemistry of Construction Materials (LC$^2$), Department of Civil and Environmental Engineering, University of California, Los Angeles, CA 90095, Email: isabella82@g.ucla.edu

[†] Laboratory for the Chemistry of Construction Materials (LC$^2$), Department of Civil and Environmental Engineering, University of California, Los Angeles, CA 90095, Email: adityaku@g.ucla.edu

[‡] Materials Science and Technology Division, Oak Ridge National Laboratory, Oak Ridge, TN 37861, Email: fieldkg@ornl.gov

[§] Physics of Amorphous and Inorganic Solids Laboratory (PARISlab), Department of Civil and Environmental Engineering, University of California, Los Angeles, CA 90095, Email: wangbu@g.ucla.edu

[\*\*] Physics of Amorphous and Inorganic Solids Laboratory (PARISlab), Department of Civil and Environmental Engineering, University of California, Los Angeles, CA 90095, Email: yuyingti@ucla.edu

[††] Fusion and Materials for Nuclear Systems Division, Oak Ridge National Laboratory, Oak Ridge, TN 37861, Email: lepapeym@ornl.gov

[‡‡] Physics of Amorphous and Inorganic Solids Laboratory (PARISlab), Department of Civil and Environmental Engineering, University of California, Los Angeles, CA 90095, Email: bauchy@ucla.edu

[§§] Corresponding Author: Laboratory for the Chemistry of Construction Materials (LC$^2$), Department of Civil and Environmental Engineering, University of California, Los Angeles, CA 90095, Email: gsant@ucla.edu

[\*\*\*] California Nanosystems Institute (CNSI), University of California, Los Angeles, CA 90095

**Significance statement**

Minerals exposed to radiation often amorphize (i.e., becoming disordered from the ordered crystalline state), which affects their properties or durability. Based on nanoscale experiments combined with atomistic simulations, we show that quartz and calcite, the two primary mineral components used in concrete, feature drastically different resistances to irradiation. The atomic network of calcite, due to its ionic/weak directional bonds, relaxes towards its pristine state upon irradiation. Quartz, however, which contains covalent/directional bonds, steadily accumulates irreversible damage. Such damage (disordering) results in an increase of its dissolution kinetics by around three orders of magnitude; seriously compromising its durability. The results indicate, in general, how a material's resistance to radiation can be improved by tuning the nature of its inter-atomic bonds.

## 1.0. Introduction and background

Concrete is used for the construction of class-I safety structures of nuclear power plants (NPPs), e.g., the containment building, biological shield and spent-fuel handling buildings, in particular. The long-term operation (LTO) of NPPs, i.e., 40+ years and beyond, is likely to be affected by concrete degradation that may modify its functionality and durability ([1],[2],[3]). Irradiation effects and alkali-silica reaction (ASR[†††]) have been identified as degradation mechanisms that require priority research in the context of license renewals and LTO of the U.S. commercial nuclear fleet ([4]). In spite of several studies ([5],[6],[7],[8],[9],[10],[11],[12],[13],[14]), comprehension of the effects of irradiation on the mechanical and physical properties of concrete remains very limited. This is because the evolution of irradiation effects depends on several factors: (i) damage can vary as a function of the neutron fluence and the γ-ray dosage, and also as a function of the concrete composition, i.e., the type of aggregate, and cement used and relative proportions of aggregates, water and cement in the mixture ([6],[13]), (ii) irradiation damage cannot be distinguished from thermal damage caused due to absorption of radiation energy ([6],[9]) and (iii) the potential impacts of irradiation, which may render the concrete vulnerable to the other physical and/or chemical degradation processes, and vice-versa.

Mineral aggregates are often composed of fragments of rocks including: granite and limestone, whose dominant constituents include quartz and calcite, respectively. Radiation damage caused to minerals such as quartz, by neutrons, or, analogously, heavy-ions has been noted to be similar, suggesting similar damage mechanisms ([15],[16]). Such damage has been described by the "direct impact model" or the "cascade model" ([17]) where interactions between radiation and minerals cause the displacement of primary knock-on atoms (PKAs). When these atoms have sufficient kinetic energy, they produce subsequent "displacements cascades" ([18]) within the remainder of the atomic network. This causes the formation of point defects that can return to their original lattice position or recombine to form other defects, e.g., defect clusters, voids, dislocations, etc. When the degree of damage or amount of energy imparted into the system reaches a threshold value, amorphization localizes and then accumulates following a homogenous or heterogeneous process. This causes the disordering[‡‡‡] of the mineral's crystal structure, resulting in a progressive loss of structural periodicity ([19]).

---

[†††] Alkali-silica reaction is caused due to the progressive dissolution of reactive silica bearing aggregates into the concrete's caustic pore-fluid (pH > 13). Following aggregate dissolution, a precipitate rich in alkaline elements, water and silica, i.e., an alkaline-silica hydrate ($Ca_x/Na_y/K_z$-$S_a$-$H_b$, where x, y, z, a, b are coefficients) of variable composition forms. This hydrate upon formation within the concrete's or aggregates porosity, eventually saturates it, resulting in deleterious cracking, and loss of mechanical properties due to internal volume expansion.

The effects of radiation damage on quartz and calcite have been described in terms of changes in physical properties, such as density, optical properties, hardness, conductivity, etc. ([20],[21],[22]). Of these, volume changes are thought to be prominent in quartz, but less significant for calcite even for neutron fluence reaching $1.0 \times 10^{20}$ $n/cm^2$ (E > 0.1 MeV) ([23],[24]). As such, the disordering or the amorphization of mineral aggregates is expected to result in two main effects:

[1] Volume changes of mineral aggregates ([25],[26],[27]) as at fixed composition, loss of crystallinity may alter the solid density, and/or,
[2] Increases in the chemical reactivity of the initially crystalline mineral aggregates ([8],[10]), as disordered materials are typically less chemically stable than their crystalline counterparts.

These effects would be problematic in the context of concrete durability as: [1] changes in volume of the aggregates can cause mechanical degradation of concretes due to microcracking of the binding cement paste, especially when the aggregate density reduces (i.e., when the aggregates expand), due to a radiation induced volumetric expansion (RIVE) ([13],[14],[27]), and [2] causes chemical degradation of the concrete due to the onset of irradiation-assisted alkali-silica reaction (IA-ASR); caused by the dissolution of siliceous aggregates into the caustic liquid-phase (pH > 13) contained in the concrete's porosity ([8],[10]). Such degradation would weaken the concrete, compromising the integrity of concrete elements that fulfill structural, shielding, and/or containment functions in NPPs. This has implications on the safety of NPPs: (a) in the event of reactor over-pressurization shock, or earthquakes, and, (b) with the increasing age of NPPs, as radiation induced degradation increases over a multi-decade service-life.

Therefore, this work elucidates how irradiation alters the atomic structures of minerals that commonly constitute aggregates in concrete. Focus is placed on contrasting how irradiation may or may not influence the chemical reactivity of a mineral with water, quantified in terms of its dissolution rate, in relation to chemical composition. The outcomes have significant implications on specifying radiation insensitive aggregates for use in NPP concretes, and for assessing the risk-profiles of which concretes (and hence nuclear power plants) may be more sensitive to radiation-induced damage than others.

## 2.0. Results and discussion

Figure 1 shows the dissolution rates of α-quartz and calcite, before and following their irradiation. Also shown are the dissolution rates of pulverized quartz, fumed silica, and of natural limestone samples for comparison. It is noted that the dissolution rates of α-quartz significantly elevate after irradiation to full amorphization; attaining near equivalence to the dissolution rate of fumed (glassy) silica for the same solvent compositions. This enhancement in the dissolution rates, by around 3 orders of magnitude indicates that the chemical stability of α-quartz is significantly compromised following irradiation. Calcite, on the other hand, shows a slight decrease in its dissolution rates following irradiation – even when the measurement uncertainty is accounted for (see Figure 2b). These results, which comprise the first direct evidence of differing irradiation-induced reactivity alterations, highlight a specificity to structure and composition – where quartz exhibits enhancements in aqueous reactivity, while calcite shows a slight reduction in reactivity following irradiation.

---

[‡‡‡] "Disordering" implies all actions which result in an increase in the entropy of a solid, at fixed composition. This includes the formation of defects and imperfections in initially crystalline mineral structures including the random movement of atoms to positions far removed from their original positions, due to ballistic (collision) effects. When ballistic effects dominate, a previously ordered, crystalline mineral, such as quartz undergoes progressive disordering until an aperiodic end-state, i.e., in terms of its atomic organization is achieved.

It should also be noted that, while minerals would show differences in their chemical reactivity in water, i.e., dissolution rates, as a function of surface orientation and surface energy, this effect is more pronounced in the case of calcite than quartz. For example, the pulverized quartz (MIN-U-SIL 10), in spite of presenting a multiplicity of surface orientations, shows a dissolution rate similar to the (001)-quartz single crystal. On the other hand, the natural limestone shows more substantial differences (±0.5 log units, see Figure 1b) in its dissolution rate, as compared to the (100)-calcite surface. While this behavior may be influenced by differences in nature of surface defects, or elemental impurities present in the calcite samples – it appears within the range of experimental uncertainty, which is higher for calcite, than for quartz.

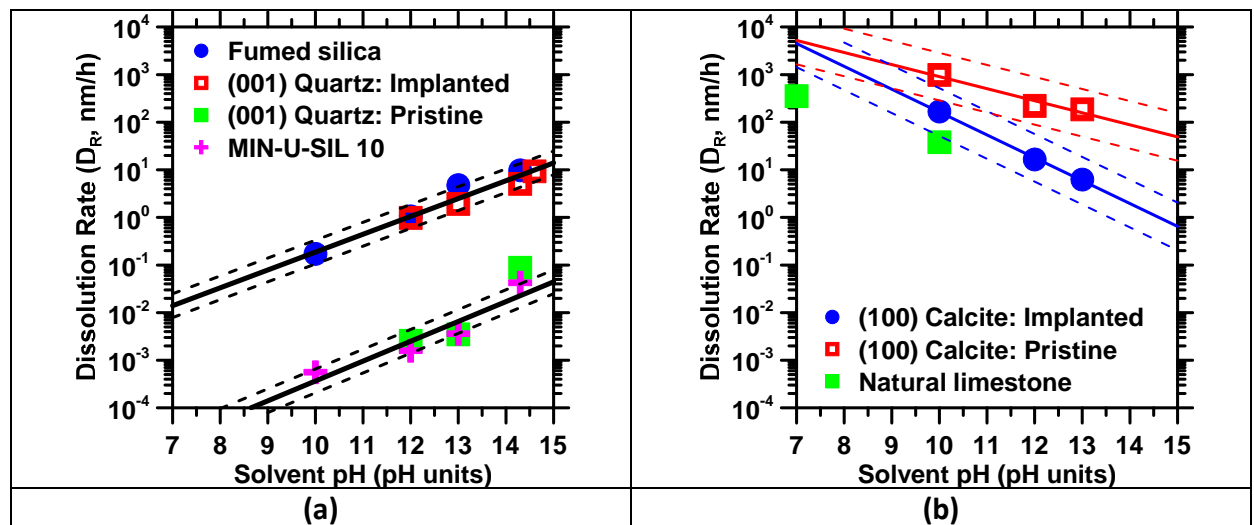

**Figure 1:** The dissolution rate (T = 25° ± 3°C, $p$ = 1 bar) as a function of solvent pH for: (a) $SiO_2$ based solids and (b) $CaCO_3$ based solids. In the case of $SiO_2$ based solids, the dissolution rate decreases with pH, while the opposite is true for $CaCO_3$ based solids. It is noted that, while quartz shows slight, if any, sensitivity to surface orientation, calcite dissolution appears more sensitive to surface orientation, and potentially solid composition (i.e., impurities present in the natural limestone). The thick solid lines show trends in dissolution rates while the thin dashed lines show the corresponding uncertainty bounds. The trend lines are fitted to an equation of the form: $D_R = A \cdot \exp(\pm B \cdot pH)$, where $A$ and $B$ are numerical constants. The highest uncertainty in the measured dissolution rate(s) is on the order of ±0.5 log units.

Near equivalence in the dissolution rates of amorphous silica, and α-quartz following irradiation suggests that the latter may be disordered under radiation exposure – a consequence which explains elevations of its dissolution rate. Dissolution has been explained within the context of crystal growth theory, focusing on the free-energy difference between the dissolving solid, and the solution in contact instead of surface speciation ([67],[70]). As such, it has been shown that dissolution occurs preferentially, and originates from high-energy sites on surfaces (structural defects and impurities), favoring the formation of etch pits (shown below). While this explanation is consistent for crystal dissolution, for compounds such as amorphous silica, the lack of atomic periodicity/structural disorder only implicates impurities in altering dissolution rates. Impurities, when present, are postulated to disrupt/weaken intermolecular bonds, destabilizing a solid, either crystalline (quartz) or amorphous (silica). Such weakening ensures that a smaller driving force is sufficient to overcome the free energy barrier; thereby making both quartz and silica more susceptible to dissolution – but, only when impurities may be present. While this view is reasonable, it does not explain, or parametrize numerical differences in the dissolution rates of compositionally analogous, and phase pure solids – e.g., quartz and amorphous silica, when they dissolve in similar solvents.

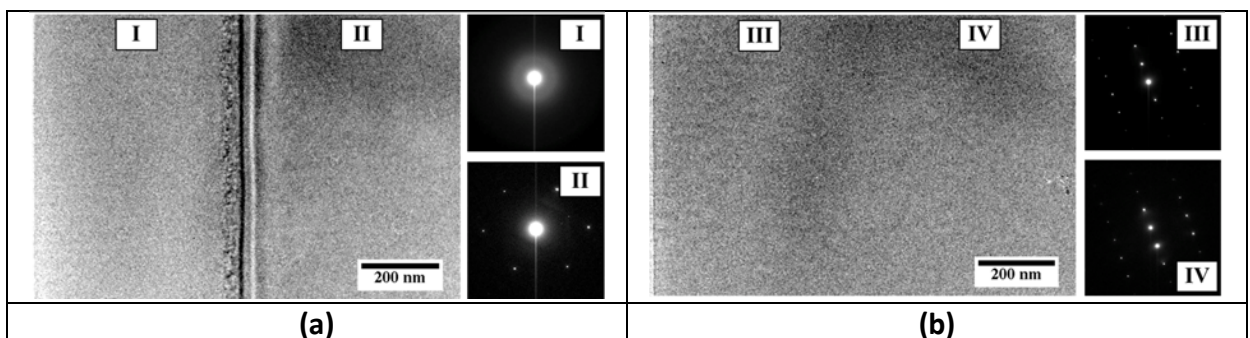

**Figure 2:** Cross-sectional TEM images and corresponding selected area electron diffraction (SAED) patterns from the ion implanted regions (I and III) and past ion end-of-range (II and IV) for: (a) α-quartz single crystal and (b) calcite single crystal. (a-I) shows the amorphization of α-quartz, while (b-III) shows the lack of amorphization in calcite. Ion implantation was carried out at an energy of 400 keV using $Ar^+$-ions for a total fluence of $1.0 \times 10^{14}$ $Ar^+/cm^2$ at room temperature. The free-sample surface is located towards the left extremity of the image(s).

To assess structural alterations induced by irradiation, cross-sectional TEM and SAED patterns were acquired and are shown in Figure 2. The sharp diffraction maxima noted in the SAED pattern past the end-of-range regions (of $Ar^+$ implantation) of quartz (Figure a-II) are not observed in SAED patterns of the corresponding ion-irradiated regions (Figure a-I). Rather, a diffuse band is noted, which is indicative of complete amorphization. This amorphous region extends ≈550 nm into the sample, in agreement with the depth of the damage zone predicted by SRIM (≈550 nm, see Figure S1). On the other hand, no changes in crystallinity are noted in calcite after irradiation – that is, SAED patterns of implanted (Figure b-IV) and pristine regions (Figure b-III) are similar and both present sharp diffraction maxima. Thus, while quartz attains a *metamict* state following ion-implantation, calcite remains resistant to radiation damage for the same implantation dose. Such composition linked structural differences were also suggested in ([13]), on the susceptibility of different minerals to irradiation. Evidence for such behavior is also noted in vibrational (FTIR) spectra acquired on irradiated samples which shows a shift toward lower wavenumbers and an intensity decrease of the asymmetric stretching modes (777, 1170 $cm^{-1}$) of quartz and of the $CO_3^{2-}$ bending/stretching modes of calcite (712, 874 and 1350 cm$^{-1}$). Such alterations in FTIR patterns have been attributed to structural modifications caused by radiation including: changes in the average Si-O-Si inter-tetrahedral angle and Si-Si bonds in quartz ([28],[29]), and the distortion and breakage of carbonate groups in calcite ([30]). These results support the idea that, while calcite is slightly, if at all, influenced by irradiation, quartz undergoes severe damage, resulting in disorder and lack of longer-range (> 10 Å) periodicity.

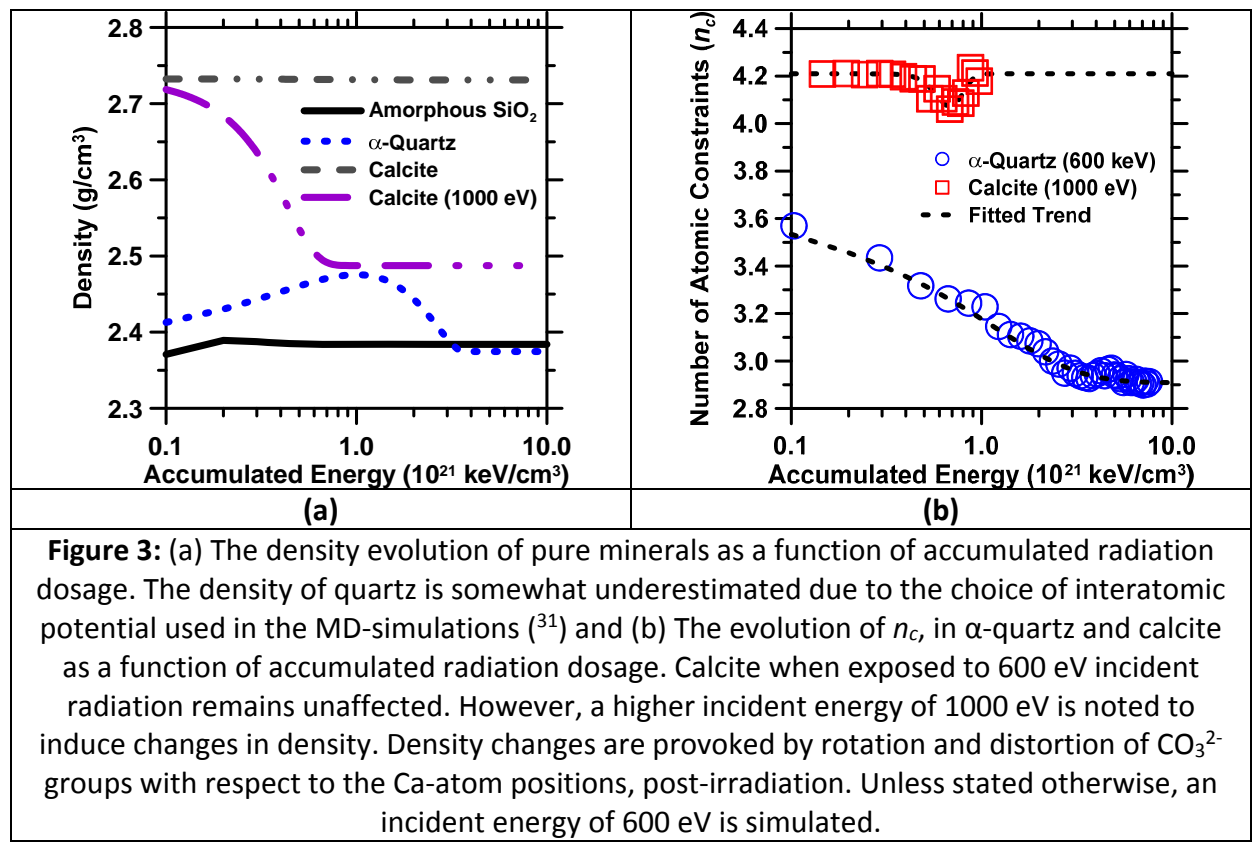

**Figure 3:** (a) The density evolution of pure minerals as a function of accumulated radiation dosage. The density of quartz is somewhat underestimated due to the choice of interatomic potential used in the MD-simulations ([31]) and (b) The evolution of $n_c$, in α-quartz and calcite as a function of accumulated radiation dosage. Calcite when exposed to 600 eV incident radiation remains unaffected. However, a higher incident energy of 1000 eV is noted to induce changes in density. Density changes are provoked by rotation and distortion of $CO_3^{2-}$ groups with respect to the Ca-atom positions, post-irradiation. Unless stated otherwise, an incident energy of 600 eV is simulated.

Molecular dynamics simulations highlight that, in agreement with the TEM-SAED analysis, that quartz is near completely disordered by radiation, while calcite is far less affected. It should be noted that the qualifier "disordered" is used in lieu of amorphized, as the resultant glassy-SiO$_2$ structure following irradiation though non-periodic, is not equivalent to amorphous silica (see Figure 2 and S2). Calcite shows substantially more resistance to radiation-induced alterations – for example, for an incident energy of 600 eV no change in the structural, or physical parameters is produced. At higher incident energies, on the order of 1000 eV, calcite experiences alterations in the form of distortion and the rotation of its $CO_3^{2-}$ groups with respect to the calcium positions. These distortions and rotations of the $CO_3^{2-}$ groups, however, alter the atomic packing density of the calcite structure – as a result of which calcite expands, resulting in a reduction in density. This expansion, which increases with the radiation dosage, achieves a limiting value, when the density of irradiated calcite stabilizes to a value of ≈90% of the pristine phase (Figure 3a). While this magnitude of expansion is larger than that estimated by Wong ([33]) following high neutron fluence exposure, given the limited data available, this difference cannot be marked to an inconsistency of the calculation scheme.

Due to the nature of interatomic potentials selected in the calculations, i.e., chosen to capture lattice dynamics accurately for disordered silica, but somewhat less so for α-quartz, the density of pristine quartz and of amorphous silica is underestimated. This underestimation is on the order of 7.5% for both the silicate solids, i.e., $\rho \approx 2.42$ g/cm$^3$ (calculated) and $\rho \approx 2.62$ g/cm$^3$ (measured) for α-quartz, and $\rho \approx 2.37$ g/cm$^3$ (calculated) and $\rho \approx 2.2$ g/cm$^3$ (measured) for disordered silica. Given that the terminal density of quartz following irradiation matches that of disordered silica, quartz would undergo a reduction in density, or conversely an increase in its molar volume of around 15%. This extent of volumetric expansion (swelling) is in excellent agreement with the analysis of Field et al. who estimated that α-quartz would swell around 14% upon its complete disordering ([13]). Amorphous silica, on the other hand, shows slight, if any changes in its density ([20]); around ≈1% across all radiation dosages (Figure 4a). It should be noted, that irradiation induces significant changes in the inter-tetrahedral (Si-O-Si) bond angles (around 7% decrease) in agreement with FTIR observations – but not the bond length, in irradiated quartz with respect to pristine quartz. As a result, a floppy, glassy disordered silica phase forms.

Structural disordering is not seen in calcite as, in general, as compared to the Si-O bond ($E_b$ = 440 kJ/mole, where $E_b$ is the bond energy) in quartz, the Ca-O bond ($E_b$ = 134 kJ/mole) in calcite is weaker, and less directional in 3D (i.e., the Ca-O-Ca bond angles show a broader distribution than Si-O-Si ones, see [58],[32] ) – as a result, under radiation induced excitations – it is free to reorganize, and show near complete recovery of initial (pristine) bond parameters once the radiation flux has ceased. It is postulated that this differing behavior is a function of the dominantly ionic character of the bonds featured in calcite, and the covalent character of quartz, an idea that was previously suggested by Wong ([33]). This suggests that radiation perturbs the weaker angular bonds, rather than stronger radial constraints: the former which, in calcite, cannot be perturbed any further. This explains why ionically bonded solids may be more resistant to radiation fluxes, than their covalent counterparts.

To comprehensively elucidate the influences of radiation on disordering the number of atomic topological constraints is computed. In solids, atoms are constrained by radial bond-stretching (BS) and angular bond-bending (BB) interactions, which act to maintain bond lengths and angles fixed around their average values. Analogous to Maxwell's stability analysis of a mechanical truss ([34]), the rigidity of a solid can be determined by enumerating the total number of constraints per atom ($n_c$, unitless), and by comparing $n_c$ to the number of degrees of freedom per atom (i.e., three in 3D). As such, atomic networks can be classified as being flexible, i.e., hypostatic, ($n_c$<3) ([35]), showing internal low-energy modes of deformation, stressed-rigid ($n_c$>3), i.e., being locked, i.e., hyperstatic, or being isostatic (i.e., statically determinate with $n_c$ = 3). As a point of note, the hardness of such atomic structures, in order of their instability scales from: flexible, to isostatic to stressed rigid networks in ascending order, i.e., from least hard to most hard ([36],[37]).

It is noted that quartz and calcite, in the pristine state both show a stressed-rigid type character. However, following irradiation, while quartz transitions to a flexible state (i.e., $n_c \approx 2.9$), calcite remains stressed-rigid (i.e., $n_c \approx 4.2$, Figure 3b). Interestingly, when the dissolution rates of these solids are cast as a function of the number of atomic constraints for a given (fixed) solvent pH – a significant trend results as shown in Figure 4(a). Specifically, in the case of $SiO_2$-based solids, the dissolution rate is noted to smoothly, and linearly increase with reducing $n_c$ – spanning from pristine to irradiated quartz, and from pulverized to fumed silica respectively. On the other hand, calcite, which shows no change in $n_c$, independent of radiation exposure, correspondingly shows little, if any change in its dissolution rate. The slight reduction that manifests in calcite dissolution rates, for irradiated calcite, is likely due to: (a) rapid dissolution of the surface exposed to solvent, such that distortions of $CO_3^{2-}$ groups, may render their removal easier, or (b) may be related to observations of a slight increase in calcite hardness, and hence stability following irradiation ([38]). If the former mechanism is operative, facilitated surface dissolution (i.e., an increase in the $CO_3^{2-}$ abundance in the solvent, in proximity to the dissolving surface) would lower the driving force for calcite dissolution faster, an effect which would slightly reduce dissolution rates (Figure 1b).

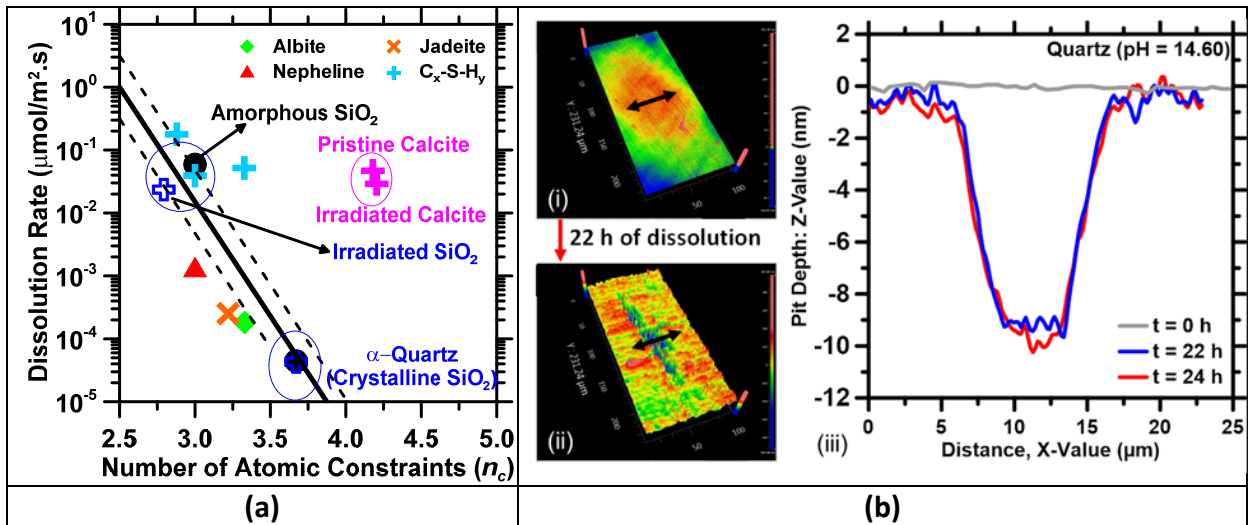

**Figure 4:** (a) A dissolution sensitivity diagram, which shows the dissolution rates of minerals and selected glasses as function of $n_c$ at pH 13; a concrete relevant pH (T = 25° ± 3°C, $p$ = 1 bar). Dissolution sensitivities of *more covalently bonded silicate minerals* increase with a decreasing number of atomic constraints ($n_c$). Decreasing $n_c$ often denotes an increase in atomic disorder, e.g., as induced by irradiation. *More ionically bonded minerals* such as calcite show little if any change in $n_c$ and dissolution rates following irradiation (Figure 1b). Rate data for albite, jadeite, and nepheline was sourced from the literature ([39]) while all other rate data was measured using VSI. The thick dashed lines show the trends in dissolution rates while the thin solid lines show relevant uncertainty bounds. The trend line is fitted to an equation of the form: $D_R = A \cdot \exp(\pm B \cdot n_c)$, where $A$ and $B$ are numerical constants. The highest uncertainty in the measured dissolution rate(s) is on the order of ±0.5 log units. (b) 3D VSI images of a (100)-surface of irradiated quartz: (i) prior to its dissolution, and (ii) and (iii) after 22 and 24 hours of dissolution at pH 14.6. The cross section profiles (ii and iii) indicates the change in the surface morphology caused due to etch pit opening/formation.

It should be noted that while Figure 4(a) shows dissolution rate correlations only for pH 13, the conclusions remains unchanged for other solution pH's – an indication of the genericity of the approach – so long as the solvent remains consistent. This diagram which indicates *the dissolution sensitivity of compositionally similar solids*, in which a specific network feature (in this case $SiO_4$ tetrahedra) controls chemical instability, captures the dissolution rate dependence on $n_c$ that is displayed by glasses compositionally analogous to albite, jadeite and nepheline ([39]), and calcium silicate hydrates (i.e., $C_x$-S-$H_y$, where C = CaO, S = $SiO_2$, and H = $H_2O$, and x and y are coefficients, where 1.2 ≤ Ca/Si ≤ 1.8, molar units), a family of disordered compounds, which comprise the primary binding and strength provisioning components of hydrated cementitious solids. These trends indicate that in regimes of high undersaturation with respect to the solute (dissolving solid), where defects are proposed to nucleate homogenously, the kinetics of the dissolution process can be characterized by $n_c$, a fundamental indicator of the chemical instability of a solid in a solvent – as a function of its atomic organization and network structure – more rigorously than the somewhat ambiguous parameter, "the degree of crystallinity".

### 3.0. Summary and Implications on concrete durability in nuclear power plants

The outcomes of this work clarify that radiation exposure, especially in the form of heavy ions, and analogously neutrons, alters the structural, physical, and chemical properties of minerals such as calcite and quartz. While the end-effects are structural (at an atomic scale), physical and chemical in the case of quartz, they are only influence the physical properties (e.g., density) of calcite. This differing behavior is correlated with the dominantly ionic nature of calcite and the covalent bonding environment in quartz, the latter of which is less resistant to radiation damage. Mineral dissolution rates are shown to be strongly correlated with a parameter known as the number of atomic constraints ($n_c$), which describes the rigidity of a network of atoms. This offers, for the first time, a quantitative means of linking the state of atomic order of a given mineral to its chemical reactivity (i.e., dissolution rate). The research demonstrates a general basis by which chemical composition-structure-property relations can be elucidated, for pristine minerals, and for minerals which have been exposed to heat, pressure or radiation, and have thus experienced irreversible alterations of their atomic structures.

This outcomes suggest different potential routes to concrete damage when calcitic or quartzitic mineral aggregates may be exposed to radiation, e.g., proximate to the reactor pressure vessel, in NPP environments. First, calcite on account of its expansion (i.e., its reduction in density, Figure 3a) is expected to induce physical damage in the concrete. Such damage will result in microcracking of the binding cement paste matrix, and the overall concrete composite, in relation to the level of (neutron) radiation exposure. However, following exposure to a threshold dosage, no further damage should evolve. Therefore, it is important to understand the implications of internal damage (microcracking) on the mechanical properties of concrete – so that its structural implications can be ascertained, and remediation measures implemented. Second, quartz due to swelling, and increasing chemical instability following irradiation, is expected to expand, causing microcracking of the cement paste matrix, and eventually dissolve in the caustic cementitious pore fluid. Since any atomic disordering is expected to be progressive, quartz is expected to dissolve incrementally faster – until a metamict state is achieved, and dissolution proceeds at a limiting-rate, i.e., of glassy silica (Figure 5a). This is problematic as the continuing dissolution of silica, so long as water, and alkali ions are available will result in the formation of an expansive alkali-silica gel, i.e., ASR[†††]. Cessation of ASR will occur only when the internal relative humidity in the concrete is lowered (RH < 80%), or when the alkalis or the siliceous aggregate are fully consumed. This is expected to need very long time scales, i.e., on the order of decades, in which time the amount of damage induced would be very substantial, and detrimental to concrete microstructure, and properties.

Both types of damage, physical and chemical are problematic as they are expected to show a gradient from the inner wall of the reactor pressure vessel (i.e., the inner surface of the reactor cavity concrete), to the outer surface of the concrete. Such gradients in expansion, i.e. strain, will result in the development of tensile hoop stresses in the reactor cavity concrete exacerbating the effects of radiation-induced damage. On a closing note, while this work has elucidated critical, and thus far unknown aspects of radiation induced alterations in minerals, significant aspects, remain worthy of evaluation: e.g., a wider range of minerals, and rocks, time dependence effects, radiation dose and energy dependence, and mechanical integrity of the affected cementitious elements. These are topics which require detailed study so that the long-term effects of radiation damage to concrete, and on NPP operations, safety and on license renewals can be evaluated.

**Acknowledgements**
The authors acknowledge full financial support for this research provided by: The Oak Ridge National Laboratory operated for the U.S. Department of Energy by UT-Battelle (LDRD Award # 4000132990), National Science Foundation (CMMI: 1066583 and 1235269), Federal Highway Administration (DTFH61-13-H-00011) and the University of California, Los Angeles (UCLA). The contents of this paper reflect the views and opinions of the authors, who are responsible for the accuracy of data presented herein. This research was conducted in: Laboratory for the Chemistry of Construction Materials (LC$^2$), the Laboratory for the Physics of AmoRphous and Inorganic Solids (PARISlab), Electron-Microscopy (EM) Core and the Molecular Instrumentation Center (MIC) at UCLA, the Michigan Ion Beam Laboratory (MIBL) and Low Activation Materials Development and Analysis (LAMDA) of Oak Ridge National Laboratory (ORNL). Support for KGF was provided by the ORNL Alvin M. Weinberg Fellowship. The authors gratefully acknowledge the support that has made these laboratories and their operations possible. This manuscript has been co-authored by UT-Battelle, LLC under Contract: DE-AC05-00OR22725 with the U.S. Department of Energy. The United States Government retains and the publisher, by accepting the article for publication, acknowledges that the United States Government retains a non-exclusive, paid-up, irrevocable, world-wide license to publish or reproduce the published form of this manuscript, or allow others to do so, for United States Government purposes.  The Department of Energy will provide public access to these results of federally sponsored research in accordance with the DOE Public Access Plan (http://energy.gov/downloads/doe-public-access-plan). The authors also acknowledge Prof. Jacob Israelachvili (UCSB) and Howard Dobbs (UCSB) for stimulating and insightful discussions on the dissolution behaviors and mechanisms of silicates.

**Materials and methods**

**Materials and ion-irradiations:** Synthetic single crystals of α-quartz and calcite with dimensions 10 mm x 10 mm x 1 mm (l x w x h) were sourced from MTI Corporation ([40]). The calcite crystals are (100)-oriented, whereas the quartz crystals are sectioned perpendicular to their optical axis (i.e., corresponding to the crystallographic *c*-axis), and are thus (001)-oriented. The quartz and calcite single-crystals were ion-beam irradiated at room temperature at the Michigan Ion Beam Laboratory (MIBL ([41])) using an implantation energy of 400 keV with $Ar^+$-ions to a total fluence of $1.0 \times 10^{14}$ ions/$cm^2$. No signs of blistering or significant sputtering were observed post-irradiation. The damage dose (dpa), the range and the concentrations of implanted ions were determined using SRIM using the quantification scheme proposed in ([42]). In addition to the single crystals, an untreated fumed silica (Cabosil HS-5), a size graded, pulverized α-quartz (MIN-U-SIL 10) and a natural limestone were also analyzed to assess their aqueous dissolution rates, so as to establish comparisons to the oriented single crystal surfaces.

**Dissolution analysis using vertical scanning interferometry:** The dissolution rates of pristine (i.e., non-implanted) and irradiated (i.e., implanted) calcite and quartz samples were measured using vertical scanning interferometry (VSI) at room temperature (25 ± 3°C). The solvents used in these studies included: reagent grade buffer solution (pH 7, 10) and NaOH solutions prepared using deionized (DI) water: 0.015 M NaOH: pH 12, 0.15M NaOH: pH 13, 2M NaOH: pH 14.3, 4M NaOH: pH 14.6. The single crystal samples were fixed on the surface of a glass slide using an inert adhesive to facilitate handling. In the case of *flat* samples, the topographical profile of the sample mapped prior to solvent contact was used as the reference plane with respect to which surface dissolution (retreat) was tracked. Powder samples were embedded in a thin-film of inert adhesive applied on the surface of a glass slide. The surface of the non-reactive adhesive once again served as the reference plane with respect to which particle dissolution was mapped ([43]).

To induce solid dissolution, a small quantity of solution (i.e., 50-to-75 μL) is applied to the sample surface using a micropipette to obtain a liquid-to-solid ratio (*l/s*, mass basis) between 50,000-to-75,000, to approximate the dilute limit. This *l/s* is appropriate to minimize the effects of solution saturation, with ions, during dissolution and limit phase precipitation, if any. After allowing for a pre-determined contact time ranging between 15-to-60 minutes (i.e., depending on the mineral dissolution rates), with reapplication of the solution if needed, the solution was removed using a compressed air stream. All measurements were carried out at ambient $pCO_2$. It should be noted that, in the manner implemented, the same solvent, of a fixed composition repetitively contacts the mineral surface. As such, no evolutions in solvent composition are permitted, and dissolution occurs at very high undersaturations with respect to the dissolving solid. As such, the dissolution rates quantified are relevant to fixed, non-evolving solvent compositions. The solution pH is the primary variable that influences the undersaturation.

A Zygo NV 9200 vertical scanning interferometer fitted with a 50x Mirau objective (N.A. = 0.55) was used in the analysis. The objective used yields a lateral spatial resolution of ≈500 nm. The interferometer has a resolution of ≈0.1 nm in the vertical, i.e., *z*-direction. The analysis scheme was organized as follows: first an image of the dry sample surface, i.e., prior to solvent contact was acquired. This constituted the "time-zero" ($t_0$) image, and dissolution was tracked using this image, and its topographical profile (i.e., of the crystal surface, or of the particles when they are embedded in adhesive) as the reference. Following solvent contact, images were periodically acquired after the removal of the solvent. These images and their topographical profiles, which were altered by dissolution, were compared to the reference image, and the change in height ($\Delta h$, nm: negative in the case of dissolution which produces surface retreat) per unit time ($\Delta t$, hours), reveals the solid's dissolution rate. Each image comprises a total scanning area of 433.81 µm x 433.81 µm in stitched mode using a 3 x 3 grid, and a back-scan length of 145 µm. The total time required for capture of the full image field is on the order of 390 seconds. It should be noted that the height reduction was mapped at up to 80 discrete points on the planar single crystal, or on particle surfaces. This statistical mapping was carried out to account for the effects of surface roughness which may differ as a function of x-y (spatial) position, and may influence dissolution rates, and to ensure that the dissolution rate quantified accounts for material inhomogeneities, if any may be present. The resulting dissolution rate ($D_R$, nm/h) is written as: $D_R = (\Delta h / \Delta t)$, where $h$ is the surface height (nm) for a given profile, and $\Delta h = h_{(i)} - h_{(i+1)}$ is the change in height between the successive steps measured over a dissolution period, $\Delta t$ (hours). It should be noted that division of $D_R$ by the molar volume ($V_M$, m$^3$/mole) of a compound reveals the dissolution rate in units of µmol/m$^2$/s. All measurements were repeated at least 3 times.

**Transmission electron microscopy:** Cross-sectional lift-outs were prepared from pristine and irradiated quartz and calcite samples using a FEI Quanta 200i DualBeam focused ion beam (FIB). Low-angle, low-energy milling was carried out following the primary thinning to obtain electron transparent sections, while minimizing any damage that may be induced by the ion-beam. The electron transparent samples were then analyzed using a Philips CM200 transmission electron microscope (TEM) at an accelerating voltage of 200 kV. Selected area electron diffraction (SAED) patterns were acquired on both pristine and irradiated regions to determine the degree, if any, of disordering in the irradiated regions. Due care was taken to minimize the electron dose to the imaged areas to reduce damage imparted by the imaging electron beam.

**Molecular dynamics simulations:** Molecular dynamics simulations were carried out on calcite and quartz structures using LAMMPS ([44]) at 300 K to study the influences of radiation damage on atomic structures ([45]). The simulated system consists of a supercell of pristine quartz or calcite containing between 4500 and 21000 atoms, depending on the incident irradiation energy. To simulate ballistic collisions induced by irradiation, a randomly selected atom is accelerated with a given kinetic energy (similar to an incident energy) to mimic energy transfer between radiation and an atom. The acceleration initiates a cascade of collisions between atoms, causing damage to the crystal structure. This process is repeated until the desired dosage is achieved. Quartz and calcite are simulated using the inter-atomic potentials of ([46]) and ([47]) respectively, which have been shown to suitably reproduce crystalline and amorphous structures of these two minerals ([45,47,48,49,50,51,52]). In order to provide a realistic prediction of high-energy events, which cause atoms to temporarily come unusually close to each other, the Ziegler, Biersack and Littmark (ZBL) potentials are used at short inter-atomic separations (< 1 Å) ([53]). Since such a small inter-atomic separation is not observed during typical conditions, i.e., close to equilibrium, the ZBL potentials only take effect during the collision cascade, while the rest of the relaxation dynamics remain unaffected. After the radiation damage simulation is completed, a series of atomic configurations are extracted at different dosage levels for detailed structural analyses. Special focus is placed on quantifying the type and nature of damage, including amorphization, and evaluating structural rigidity ([54,55]) by analysis of atomic trajectories. The rigidity analysis consists of an enumeration of the number of intact and broken radial and angular bond constraints at 300 K ([56,57]). A bond constraint is considered to be broken if the relative variation in the bond distance (or bond angle) is sufficiently large, i.e., exceeding 7% ([58]), which indicates the absence of an underlying restoring force that would maintain the bond length (and angle) fixed around its average value. It should be noted that the constraints enumeration procedure does not significantly depend on the choice of this threshold, similar to the Lindemann criterion ([58]).

**Supplementary Information**

- **The influence of solution composition on quartz and calcite dissolution rates**

The dissolution rate of quartz increases with pH that of calcite decays. This pH dependence of quartz dissolution has been previously explained in terms of: the undersaturation of the bulk solution, the speciation of silicon, and surface complexations of species ([59,60,61,62,63]). For pH > 6, negatively charged silicate species ($H_2SiO_4^{2-}$, $H_3SiO_4^-$) are present in solution and protonated "≡SiO-Na$^+$" and de-protonated "≡SiO$^-$" species are present on solid silicate surfaces. The latter species are thought to be amenable to rapid silica dissolution. It has been argued that when alkali cations, e.g., Na$^+$ or K$^+$, are present in the bulk solution, they cause a redistribution of surface complexes, such that ≡SiO-Na$^+$ or ≡SiO-K$^+$ become more dominant than ≡SiO$^-$ species, further enhancing quartz dissolution ([59,60,61,62,63,64,65,66,67]). Such increases in silicate dissolution rates in the presence of alkaline salts, e.g., NaCl, KCl, CaCl$_2$, MgCl$_2$, is termed as the "salt effect" ([59,67,68,69,70]), wherein the reactivity enhancement induced by alkaline cations follows an order similar to a Hoffmeister series where: $Ba^{2+}$ > $Sr^{2+}$ > $K^+$ > $Na^+$ > $Mg^{2+}$ > $Li^+$ for amorphous silica and $Ba^{2+}$ > $K^+$ ≈ $Na^+$ ≈ $Li^+$ > $Ca^{2+}$ > $Mg^{2+}$ for α-quartz ([61,64]).

Calcite dissolution at low pH is proportional to the [H$^+$ and CO$_3^{2-}$] abundance and is impacted by mass transfer (i.e., diffusion of ions from the solid surface into the bulk solution) ([71,72,73,74,75]). On the other hand, at neutral to alkaline conditions, calcite dissolution is pH-independent and controlled by the presence of high energy sites on the surface, i.e., interface control ([76,77,78,79]). The presence of these sites is revealed by the formation of etch pits during dissolution. Pits nucleate rapidly far from solute equilibrium. As such, while 2D pits nucleate spontaneously even in defect-free areas at very high undersaturations, defect-assisted nucleation (i.e., that which is associated with structural/impurity defects) operates at intermediate undersaturation levels, with progressive step-retreats controlling dissolution rates nearer to equilibrium ([80]). On {10$\bar{1}$4} surfaces, long-lived, larger and pointed pits have been attributed to the presence of line defects, while short-lived, smaller and flat-bottom pits have been attributed to point defects ([77,81]). These pits show a rhombic morphology ([72,77,79,81,82,83,84]), though pyramidal pits may also form on {10$\bar{1}$0} surfaces ([85]), as is seen in the present study (not shown). Similar to silica-based minerals, the dissolution of calcite is also affected by the presence of divalent cations within the solid ([79]). At neutral to basic pH, $Ca^{2+}$, $Mg^{2+}$, $Sr^{2+}$ and $Ba^{2+}$ ions and metals such as Cd, Ni, Cu, Co and Mn inhibit calcite dissolution due to interactions with its surface ([83,84,86,87,88,89,90,91,92,93,94]).

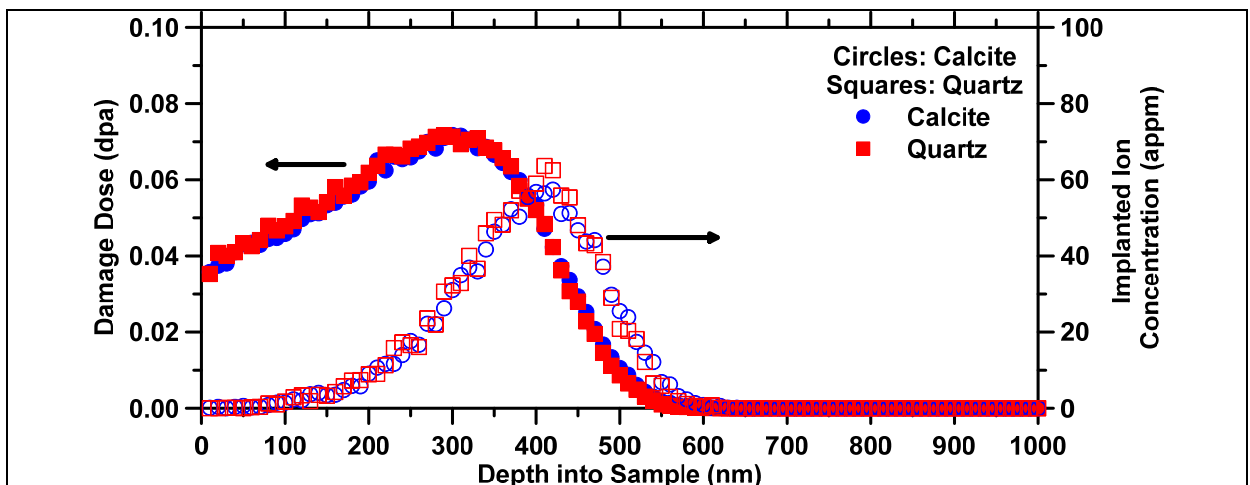

**Figure S1:** The calculated damage dose (dpa) and implanted ion concentration (appm) as a function of depth into the sample for calcite and α-quartz. These values are calculated using the scheme implemented in ([95]) for an implantation energy of 400 keV using $Ar^+$-ions for a total fluence of $1.0 \times 10^{14}$ $Ar^+/cm^2$. Ion implantation was carried out at room temperature and the free-sample surface is located at "x = 0 nm" – towards the left extremity of the plot. As a function of their similar density pre-irradiation, $Ar^+$ ions are implanted to a depth of ≈550 nm in both mineral samples.

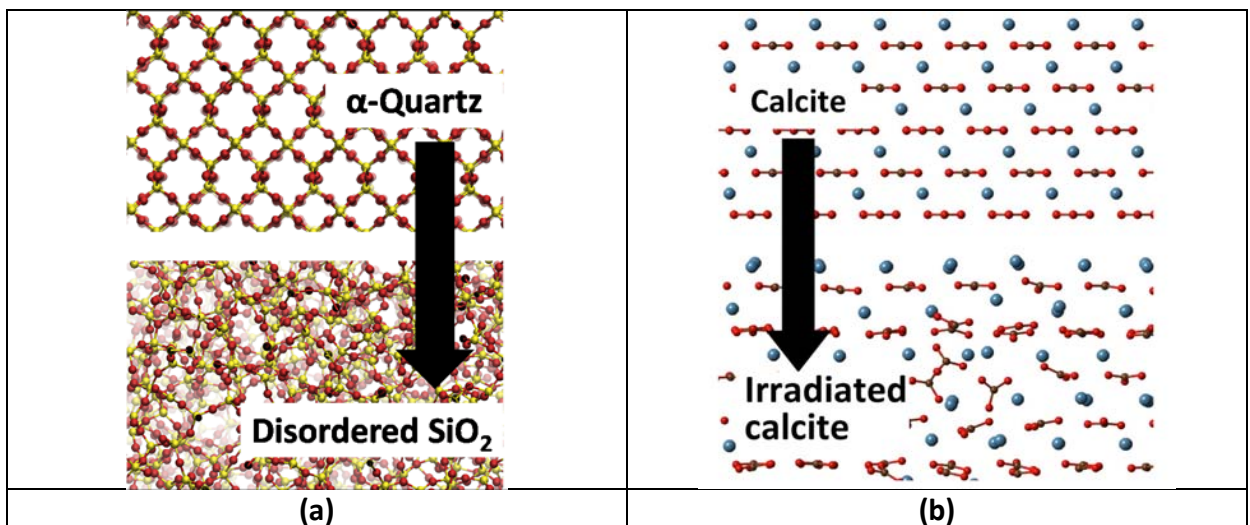

**Figure S2:** (a) The progressive disordering of α-quartz to disordered silica under energetic particle exposure (Left image: red: oxygen atoms, yellow: silicon atoms) and (b) The evolution of the atomic structure of calcite following irradiation. Structural evolution in calcite manifests only in the form of the carbonate groups experiencing random rotations and distortion in the irradiated structure (Right image: blue: calcium atoms, red: oxygen atoms, brown: carbon atoms). The structural data were sourced from molecular dynamics (MD) simulations and visualized using VMD ([96]).

**References**

[1] Neville A (1963) Properties of concrete. John Wiley and Sons Ltd, 844 pp.

[2] Mehta P, Monteiro P (2006) Concrete-Microstructure, Properties and Materials. 3rd ed. McGraw-Hill Professional, New York, 653 pp.

[3] Naus D (2007) Primer on Durability of Nuclear Power Plant Reinforced Concrete Structures - A Review of Pertinent Factors. Technical Report *NUREG/CR-6927 - ORNL/TM-2006/529, Oak Ridge National Laboratory - U.S. Nuclear Regulatory Commission.*

[4] Graves H, Le Pape Y, Naus D, Rashid J, Saouma V, Sheikh A, Wall J (2013) Expanded material degradation assessment (EMDA). Volume 4. Aginh of concrete. Technical Report NUREG/CR-7153, ORNL/TM-2011/545. U.S. Nuclear Regulatory Commission.

[5] Clark R. (1958) Radiation damage to concrete. Technical Report HW-56195. General Electric. Hanford Laboratories, Richland, WA.

[6] Hilsdorf HK, Kropp J, Koch HJ (1978) The effects of nuclear radiation on the mechanical properties of concrete. *ACI Spec Publ* 55:223–251.

[7] Naus D (1986) Concrete Component Aging and its Significance Relative to Life Extension of Nuclear Power Plant. Technical Report *NUREG/CR-6927 - ORNL/TM-2006/5298*, *Oak Ridge National Laboratory - U.S. Nuclear Regulatory Commission*.

[8] Ichikawa T, Koizumi H (2002) Possibility of radiation-induced degradation of concrete by alkali-silica reaction of aggregates. *J Nucl Sci Technol* 39:880-884.

[9] Fillmore DL (2004) Literature Review of the Effects of Radiation and Temperature on the Aging of Concrete. Technical Report INEL/EXT-04-02319. Idaho National Engineering and Environmental Laboratory.

[10] Ichikawa T, Kimura T (2007) Effect of Nuclear Radiation on Alkali-Silica Reaction of Concrete. *J Nucl Sci Technol* 44:1281–1284.

[11] Fujiwara K, Ito M, Sasanuma M, Tanaka H, Hirotani K, Onizawa K, Suzuki M, Amezawa H (2009) Experimental Study of the Effect of Radiation Exposure to Concrete, *Proceedings of the 20th International Conference on Structural Mechanics in Reactor Technology, SMiRT 20 - Division I.*

[12] Kontani O, Sawada S, Maruyama I, Takizawa M, Sato O (2013) Evaluation of Irradiation Effects on Concrete Structure: Gamma-Ray Irradiation Tests on Cement Paste. *ASME 2013 Power Conference* (American Society of Mechanical Engineers), 1-8 pp.

[13] Field KG, Remec I, Le Pape Y (2015) Radiation effects in concrete for nuclear power plants - Part I: Quantification of radiation exposure and radiation effects. *Nucl Eng Des* 282:126-143.

[14] Le Pape Y, Field KG, Remec I (2015) Radiation effects in concrete for nuclear power plants. Part II: Perspective from micromechanical modelling. *Nucl Eng Des* 282:144-157.

[15] Douillard L, Durand JP (1996) Amorphization of α-quartz under irradiation. *J Phys III France* 6:1677-1687.

[16] Douillard L, Durand JP (1996) Swift heavy ion amorphisation of quartz - a comparative study of the particle amorphization mechanism of quartz. *Nucl Instr and Meth in Phys Res B* 107:212-217.

[17] Ewing RC, Meldrum A, Wang L, Wang S (2000) Radiation-Induced Amorphization. *Rev Mineral Geochem* 39(1):319–361.

[18] Kinchin GH, Pease RS The displacement of atoms in solids by radiation. *Rep Prog Phys* 18:1–51.

[19] Okamoto PR, Lam NQ, Rehn LE (1999) Physics of crystal-to-glass transformations. *Solid State Phys* 52:2-133.

[20] Primak W (1958) Fast-neutron-induced changes in quartz and vitreous silica. *Phys Rev* 110:1240.

[21] Raju KS (1977) Radiation effects in calcite. *Pramāṇa* 8(3):266-275.

[22] Rao EV, Murthy MR (2008) Ion beam modifications of defect sub-structure of calcite cleavages. *Bull Mater Sci* 31:139-142.

[23] Hauser O, Schenk M (1964) Veränderungen der Kristallstruktur einiger Oxyde, Karbonate und Titanate durch Neutronenbestrahlung. *Phys Status Solidi B* 6:83-88.

[24] Luthra J (1969) X-ray studies on pile-irradiated calcite. *Indian J Pure Ap Phy* 7:444-445.

[25] Wittels M (1957) Structural behavior of neutron irradiated quartz. *Philos Mag* 2:1445-1461.

[26] Zubov V, Ivanov A (1966) Expansion of quartz caused by irradiation with fast neutrons. *Sov Phys Crystallogr* 11:372-374.

[27] Seeberger J, Hilsdorf H (1982) Einfluss von radioactiver Strahlung auf die Festogkeit and Struktur von beton. Technical Report NR2005. Institut für Massivbau and Baustofftechnologie, Universität Karlsruhe.

[28] Mazzoldi P, Carnera A, Caccavale F, Favaro ML (1991) N and Ar ion-implantation effects in $SiO_2$ films on Si single-crystal substrates. *J Appl Phys* 70:3528-3536.

[29] Battaglin G, Boscolo-Boscoletto A, Caccavale F, De Marchi G, Mazzoldi P, Arnold GW (1992) Etching effects in ion Implanted $SiO_2$. *Modifications Induced by Irradiation in Glasses* (Elsevier Science Publishers B. V., Amsterdam), pp 91–96.

[30] Nagabhushana H, Prashantha SC, Nagabhushana BM, Lakshminarasappa BN, Singh F (2008) Damage creation in swift heavy ion-irradiated calcite single crystals: Raman and Infrared study. *Spectrochim Acta A Mol Biomol Spectrosc* 71:1070–1073.

[31] Campañá C, Müser MH, Tse JS, Herzbach D, Schöffel P (2004) Irreversibility of the pressure-induced phase transition of quartz and the relation between three hypothetical post-quartz phases. *Phys Rev B* 70:224101.

[32] Bauchy M (2012) Structural, vibrational, and thermal properties of densified silicates: Insights from molecular dynamics. *J Chem Phys* 137:044510.

[33] Wong, C. (1974) Neutron radiation damage in some birefringent crystals. *Physics Letters A*, 1974, *50*, 346.

[34] Maxwell JC (1864) L. On the calculation of the equilibrium and stiffness of frames. *Philos. Mag.* 27:294-299.

[35] Cai Y, Thorpe MF (1989) Floppy modes in network glasses. *Phys Rev B* 40:10535-10542.

[36] Bauchy M, Qomi MJA, Bichara C, Ulm F-J, Pellenq RJM (2015) Rigidity Transition in Materials: Hardness is Driven by Weak Atomic Constraints. *Phys Rev Lett* 114:125502.

[37] Smedskjaer MM, Mauro JC, Yue Y (2010) Prediction of Glass Hardness Using Temperature-Dependent Constraint Theory. *Phys. Rev. Lett.* 105:115503.

[38] Raju KS (1977) Radiation effects in calcite. *Pramana* 8(3): 266-275.

[39] Hamilton JP, Brantley SL, Pantano CG, Criscenti LJ, Kubicki JD (2001) Dissolution of nepheline, jadeite and albite glasses: toward better models for aluminosilicate dissolution. *Geochim Cosmochim Acta* 65:3683-3702.

[40] http://www.mtixtl.com/, last accessed, (April 2015)

[41] Rotberg VH, Toader O, Was GS (2001) A High Intensity Radiation Effects Facility in: *Applications of Accelerators in Research and Industry, edited by J. L. Duggan and I. L. Morgan, AIP Conf. Proc. 578,* American Institute of Physics, Melville, NY, 687-691

[42] Stoller RE, et al. (2013) On the use of SRIM for computing radiation damage exposure. *Nucl Instrum Methods Phys Res Sect B Beam Interact Mater At* 310:75–80.

[43] Kumar, A, Reed J, Sant G (2013) Vertical Scanning Interferometry: A New Method to Measure the Dissolution Dynamics of Cementitious Minerals. *Jour. Amer. Cer. Soc.* 96(9): 2766-2778

[44] Plimpton S (1995) Fast parallel algorithms for short-range molecular dynamics. *J. Comput. Phys.* 117:1–19.

[45] Wang B, Yu Y, Pignatelli I, Sant G, Bauchy M (2015) Nature of Radiation-Induced Defects in Quartz. *arXiv preprint arXiv:1504.02537*.

[46] van Beest BWH, Kramer GJ, van Santen RA (1990) Force fields for silicas and aluminophosphates based on ab initio calculations. *Phys Rev Lett* 64(16):1955-1958.

[47] Vuilleumier R, Seitsonen A, Sator N, Guillot B (2014) Structure, equation of state and transport properties of molten calcium carbonate (CaCO3) by atomistic simulations. *Geochim Cosmochim Acta* 141:547-566.

[48] Phillips CL, Magyar RJ, Crozier PS (2010) A two-temperature model of radiation damage in α-quartz. *J. Chem Phys* 133(14): 144711-144721.

[49] Wang B, Yu Y, Lee YJ, Bauchy M (2015) Intrinsic Nano-Ductility of Glasses: The Critical Role of Composition *Front Mater* 2.

[50] Yuan F, Huang L (2014) Brittle to Ductile Transition in Densified Silica Glass. *Scientific Reports* 4.

[51] Sifré D, Hashim L, Gaillard F (2014) Effects of temperature, pressure and chemical compositions on the electrical conductivity of carbonated melts and its relationship with viscosity. *Chem Geol*.

[52] Vuilleumier R, Seitsonen AP, Sator N, Guillot B (2015) Carbon dioxide in silicate melts at upper mantle conditions: Insights from atomistic simulations. *Chem Geol*.

[53] Ziegler JF, Biersack JP, Littmark U, Anderson HH (1985) *The stopping and ranges of ions in matter. Vol 1* (Pergamon, New York).

[54] Mauro JC (2011) Topological constraint theory of glass. *Am Ceram Soc Bull* 90:31-37.

[55] Phillips JC (1979) Topology of covalent non-crystalline solids I: Short-range order in chalcogenide alloys. *J Non-Cryst Solids* 34:153-181.

[56] Bauchy M, Micoulaut M (2011) Atomic scale foundation of temperature-dependent bonding constraints in network glasses and liquids. *J Non-Cryst Solids* 357:2530-2537.

[57] Bauchy M, Micoulaut M, Celino M, Le Roux S, Boero M, Massobrio C (2011) Angular rigidity in tetrahedral network glasses with changing composition. *Phys Rev B* 84:054201.

[58] Bauchy M, Abdolhosseini Qomi MJ, Bichara C, Ulm FJ, Pellenq RJM (2014) Nanoscale Structure of Cement: Viewpoint of Rigidity Theory. *J Phys Chem C* 118:12485-12493.

[59] Dove PM, Crear DA (1990) Kinetics of quartz dissolution in electrolyte solutions using a hydrothermal mixed flow reactor. *Geochim Cosmochim Acta* 54:955-969.

[60] Seward TM (1974) Determination of the first ionization constant of silicic acid from quartz solubility in borate buffer solutions to 350°C. *Geochim Cosmochim Acta* 38:1651–1664.

[61] Plettinck S, Chou L, Wollast R (1994) Kinetics and mechanism of dissolution of silica at room temperature and pressure. *Mineral Mag*, 58: 728-729.

[62] House WA (1994) The role of surface complexation in the dissolution kinetics of silica: effects of monovalent and divalent ions at 25°C. *J Colloid Interface Sci* 163:379-390.

[63] Dove PM (1994) The dissolution kinetics of quartz in sodium chloride solutions at 25°C to 300°C. *Am J Sci* 294:665-712.

[64] Dove PM, Nix CJ (1997) The influence of the alkaline earth cations, magnesium, calcium and barium on the dissolution kinetics of quartz. *Geochim Cosmochim Acta* 61:3329–3340.

[65] Dove PM (1999) The dissolution kinetics of quartz in aqueous mixed cation solutions. *Geochim Cosmochim Acta* 63 (22):3715–3727.

[66] Icenhower JP, Dove PM (2000) The dissolution kinetics of amorphous silica into sodium chloride solutions: effects of temperature and ionic strength. *Geochim Cosmochim Acta* 64:4193–4203.

[67] Dove PM, Han N, De Yoreo JJ (2005) Mechanisms of classical crystal growth theory explain quartz and silicate dissolution behavior. *Proc Natl Acad Sci U S A* 102(43):15357–15362.

[68] Diénert F, Wandenbulke F (1923) Sur le dosage de la silice dans les eaux. *Comptes Rendus Séances Académie Sci* 176:1478–1480.

[69] van Lier JA, de Bruyn PL, Overbeek TG (1960) The solubility of quartz. *J Phys Chem* 64:1675-1682.

[70] Dove PM, Han N, Wallace AF, De Yoreo JJ (2008) Kinetics of amorphous silica dissolution and the paradox of the silica polymorphs. *Proc Natl Acad Sci* 105(29):9903–9908.

[71] Rickard D, Sjöberg EL (1983) Mixed kinetic control of calcite dissolution rates. *Am J Sci* 283:815-830.

[72] Shiraki R, Rock PA, Casey WH (2000) Dissolution kinetics of calcite in 0.1 M NaCl solution at room temperature: an atomic force microscopic (AFM) study. *Aquat Geochem* 6:87–108.

[73] Sjöberg EL, Rickard DT (1984) Calcite dissolution kinetics: surface speciation and the origin of the variable pH dependence. *Chem Geol* 42:119-136.

[74] Sjöberg EL, Richard DT (1985) The effect of added dissolved calcium on calcite dissolution kinetics in aqueous solutions at 25°C. *Chem geol* 49:405-413.

[75] Compton RG, Pritchard KL (1990) The dissolution of calcite at pH> 7: kinetics and mechanism. *Philos Trans R Soc Lond Ser Math Phys Sci* 330(1609):47–70.

[76] Schott J, Brantley S, Crear D, Guy C, Borcsik M, Willaime C (1989) Dissolution kinetics of strained calcite. *Geochim Cosmochim Acta* 53:373–382.

[77] MacInnis IN, Brantley SL (1992) The role of dislocations and surface morphology in calcite dissolution. *Geochim Cosmochim Acta* 56:1113–1126.

[78] VanCappellen P, Charlet L, Stumm W, Wersin P (1993) A surface complexation model of the carbonate mineral-aqueous solution interface. *Geochim Cosmochim Acta* 57:3505–3518.

[79] Atanassova R, Cama J, Soler JM, Offeddu FG, Queralt I, Casanova I (2013) Calcite interaction with acidic sulphate solutions: a vertical scanning interferometry and energy-dispersive XRF study. *Eur J Mineral* 25:331–351.

[80] Teng HH (2004) Controls by saturation state on etch pit formation during calcite dissolution. *Geochim Cosmochim Acta* 68:253–262.

[81] Liang Y, Baer DR, McCoy JM, Amonette JE, LaFemina JP (1996) Dissolution kinetics at the calcite-water interface. *Geochim Cosmochim Acta* 60:4883–4887.

[82] Duckworth OW, Martin ST (2004) Dissolution rates and pits morphologies of rhombohedral carbonate minerals. *Am Mineral* 89:554-563.

[83] Vinson MD, Arvidson RS, Lüttge A (2007) Kinetic inhibition of calcite (104) dissolution by aqueous manganese (II). *J Cryst Growth* 307:116-125.

[84] Xu and Higgins, 2001; Xu M, Higgins SR (2011) Effects of magnesium ions on near-equilibrium calcite dissolution: step kinetics and morphology. *Geochim Cosmochim Acta* 75:719–733.

[85] Thomas JM, Renshaw GD (1965) Dislocations in calcite and some of their chemical consequences. *Trans Faraday Soc* 61:791–796.

[86] Morse JW, Arvidson RS (2002) The dissolution kinetics of major sedimentary carbonate minerals. *Earth-Sci Rev* 58:51-84.

[87] Ardvinson RS, Collier M. Davis KJ, Vinson MD (2006) Magnesium inhibition of calcite dissolution kinetics. *Geochim Cosmochim Acta* 70:583-594.

[88] Lea AS, Amonette JE, Baer DR, Liang Y, Colton NG (2001) Microscopic effects of carbonate, manganese and strontium ions in calcite dissolution. *Geochim Cosmochim Acta* 65:369-379.

[89] Sjöberg EL (1978) Kinetics and mechanism of calcite dissolution in aqueous solutions at low temperatures. *Stockholm Contrib Geol* 32: 32 pp.

[90] Buhmann D, Dreybrodt W (1987) Calcite dissolution kinetics in the system $H_2O$-$CO_2$-$CaCO_3$ with participation of foreign ions. *Chem Geol* 64:89-102.

[91] Gytjahr A, Dabringhaus H, Lacmann R (1996) Studies of the growth and dissolution kinetics of the $CaCO_3$ polymorphs calcite and aragonite: II the influence of divalent cation additives on the growth and dissolution rates. *J Cryst Growth* 158:310-315.

[92] Salem MR, Mangood AH, Hamdona SK (1994) Dissolution ofcalcite crystals in the presence of some metal ions. *J Mater Sci* 29:6463-6467.

[93] Martin-Garin A, Van Cappellen P, Charlet L (2003) Aqueous cadmium uptake by calcite: a stirred flow-through reactor study. *Geochim Cosmochim Acta* 67:2763-2774.

[94] Plummer LN, Mackenzie FT (1974) Predicting mineral solubility from rate data – application to the dissolution of magnesium calcites. *Am J Sci* 274:61-83.

[95] Stoller RE, Toloczko MB, Was GS, Certain AG, Dwaraknath S, Garner FA (2013) On the use of SRIM for computing radiation damage exposure. Nuclear Instruments and Methods in Physics Research Section B: Beam Interactions with Materials and Atoms 310: 75–80.

[96] Humphrey W, Dalke A, Schulten K (1996) VMD: visual molecular dynamics. *J Mol Graphics* 14:33-38.